\documentclass[sn-mathphys-num]{sn-jnl}

\usepackage{graphicx}%
\usepackage{multirow}%
\usepackage{amsmath,amssymb,amsfonts}%
\usepackage{amsthm}%
\usepackage{mathrsfs}%
\usepackage[title]{appendix}%
\usepackage{xcolor}%
\usepackage{textcomp}%
\usepackage{manyfoot}%
\usepackage{booktabs}%
\usepackage{algorithm}%
\usepackage{algorithmicx}%
\usepackage{algpseudocode}%
\usepackage{listings}%

\theoremstyle{thmstyleone}%
\theoremstyle{thmstyletwo}%

\theoremstyle{thmstylethree}%

\begin{document}

\title[Article Title]{Square-well model for superconducting pair-potential}


\author{\fnm{Erkki} \sur{Thuneberg}}



\affil{\orgdiv{QTF Centre of Excellence}, \orgname{Aalto University}, \orgaddress{\city{Espoo}, \postcode{FI-00076}, 
\country{Finland}}}




\abstract{We study  Andreev reflection in a one-dimensional square-well pair-potential. We discuss the history of the model. The current-phase relation is presented as a sum over Matsubara frequencies. How the current arises from bound and continuum levels is found by analytic continuation. We discuss two limiting cases of the square-well potential, the zero-length well and the infinite well. The model is quantitatively valid in some cases but forms the basis for understanding a wide range of problems in inhomogeneous superconductivity and superfluidity.
}

\keywords{Andreev bound state, SNS junction, square-well pair-potential, Josephson current}



\maketitle
Andreev reflection is a fundamental process in superconductivity. The process consists of a particle type excitation being reflected as hole type excitation. In contrast to usual reflection, the momentum change in Andreev reflection is small compared to the particle's momentum, as the hole type excitation moves  in the opposite direction than its momentum.  Andreev reflection treats the transport of electricity and heat differently. It is the essential process that allows electric current to flow in any inhomogeneous superconductor but it resists the heat current. 

\begin{figure}[htb] 
   \centering
   \includegraphics[width=0.55\linewidth]{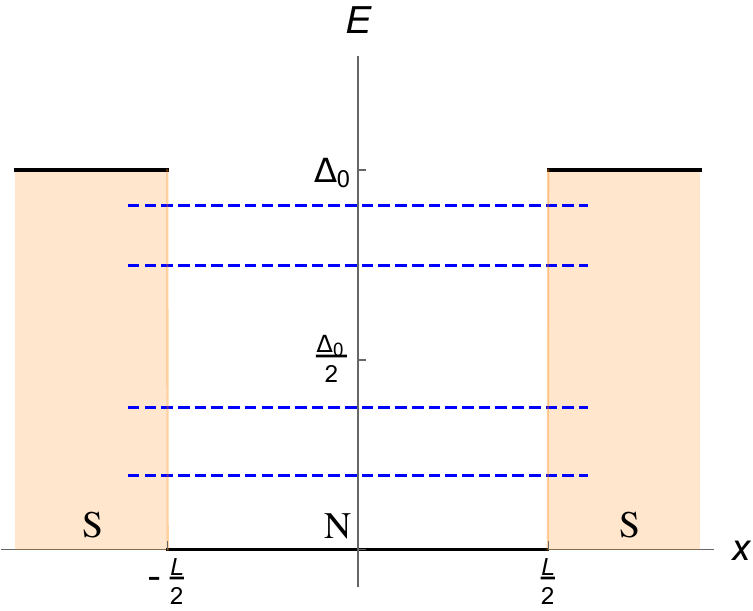}     
   \caption{The square-well potential for the pair potential $\Delta(x)$: it equals $\Delta_0 e^{-{\rm i}\phi/2}$ for $x<-L/2$, $0$ for $|x|<L/2$, and $\Delta_0 e^{{\rm i}\phi/2}$ for $x>L/2$. The figure also shows bound state energies (dashed lines). These are evaluated with parameter values $L\Delta/\hbar v_F=4.5$ and $\phi=1$ rad. At energies $E>\Delta_0$ the states form a continuum.  }
   \label{f.pot}
\end{figure}

The purpose of this article is to study Andreev reflection in a simple model. We model the pair potential by a {\em square-well}  form, see Fig.\ \ref{f.pot}. This could be compared with the model of {\em square-well potential }  and its special case the {\em infinite potential well}, that are studied in practically all textbooks of quantum mechanics. The study of such  models has a twofold justification. 1) Although the model is not precisely realized in Nature, it still forms a reasonable approximation for several applications. 2) A model that can be solved precisely allows one to learn physics, which is useful also in cases where the model ceases to be correct quantitatively.

We specify the parameters of the square-well model (Fig.\ \ref{f.pot}). We consider the pair potential $\Delta(x)$ that depends on a single coordinate $x$. The potential is assumed to vanish in the well of width $L$. Outside of this range, the amplitude is constant, $|\Delta|=\Delta_0$. We allow a phase difference $\phi$ between the two sides. Because the absolute phase is not important, it is convenient to choose the phase symmetrically, $\Delta(x)=\Delta_0 e^{-{\rm i}\phi/2}$ for $x<-L/2$,  and $\Delta=\Delta_0 e^{{\rm i}\phi/2}$ for $x>L/2$. 
It is convenient to call the middle region as normal (N), and the surrounding regions as superconducting banks  (S).
We can think that $x$ is a coordinate along a quasiparticle trajectory in a bulk metal. Alternatively, we can consider the system to represent a single quantum channel. (For an introduction to quantum channels see Ref.\ \cite{Datta}.) We neglect all scattering (by interfaces, walls, impurities or lattice vibrations) except the one caused by the pair potential. We count the energies starting from the Fermi level. For brevity, the bank gap amplitude $\Delta_0$ is denoted by $\Delta$ in the following. 
 
The motivation for Andreev was to understand experimental observations that the heat conductivity in the intermediate state was found much smaller than in the normal or zero-field superconducting state. The intermediate state of type I superconductors, which occurs in the magnetic field, was known to have a lamellar structure consisting of alternating normal and superconducting regions. This naturally leads to thinking that the normal-superconducting interfaces could be responsible for the reduced heat conductivity. The problem, however, is that the force on the electron at such an interface is too weak to cause ordinary scattering of electrons that have momentum near the Fermi momentum. Andreev solved this puzzle by inventing a new type of reflection, where an incident electron is reflected as a hole-type excitation. The latter travels in the opposite direction at a momentum that is only slightly changed from that of the incident electron. In the first publication on the topic \cite{Andreev64}, he used Gor'kov's equations to solve the reflection problem for a pair-potential step. More generally, he showed that the result for a realistic N-S interface is not essentially different for electron energies slightly above the gap, the relevant ones for heat conduction at a low temperature. This allowed him to estimate the heat conductivity of the electrons at low temperature, in good agreement with experiment. 

Andreev studied the problem of the square-well potential (Fig.\ \ref{f.pot}) in Ref.\ \cite{Andreev65}. At energies below $\Delta$, an electron is reflected from one step as a hole, which is then reflected back from the other step as an electron. This leads to the formation of discrete energy levels, which we now call {\em Andreev bound states}. For low-energy levels, he found equally spaced energy values
\begin{eqnarray}
E_{n}=\frac{\pi\hbar v_{F}}{L}(n+\gamma).
\label{e.kedr}\end{eqnarray}
Here $v_F$ is the Fermi velocity, $n$ is an integer and  $\gamma$ is a constant, and it is assumed $L\gg \hbar v_F/\Delta$. Note that the level spacing is the same as one would obtain in a deep ordinary potential well at energies close to the Fermi energy.  Andreev applied this to the  intermediate state, where the discreteness of the levels should be visible in thermodynamic properties at temperatures $T\lesssim\hbar  v_F/L$.

The next important development was by Kulik \cite{Kulik69}. He had the idea that the square-well pair-potential could be used as a model of the SNS Josephson junction. For that, he took into account that there could be a phase difference between the superconducting banks. For the bound state energies, he found the equation
\begin{eqnarray}
E_{n,\pm}=\frac{\hbar v_{F}}{2L}\left(\pm \phi+2\pi n+2\arccos\frac{E_{n,\pm}}{\Delta}\right).
\label{e.kedr2}\end{eqnarray}
This formula is valid for all bound state energies and arbitrary $L$. For low-energy levles, the last term in the parenthesis approaches $\pi$. Thus the energy (\ref{e.kedr2}) reduces to the Andreev's result (\ref{e.kedr}) but now gives a 
phase-dependent $\gamma(\phi)= \frac12 \pm\phi/2\pi$. The levels that increase in energy with increasing $\phi$ correspond to momentum to the right. This means that there is an electron moving to the right and a hole to the left. For levels that decrease in energy with increasing $\phi$, it is the opposite. Thus, depending on the phase difference and the occupations of these levels, there is a current through the system. Thus the SNS system forms a Josephson junction. 

Kulik went on to calculate the current. His method was complicated. In particular, the contributions from the bound states and continuum were calculated separately. The result reported in Ref.\ \cite{Kulik69} turned out to be incorrect. In particular, the continuum contribution was found to have a phase dependence of $\sin\phi$, contrary to later calculations to be discussed below.

The Josephson current was next calculated by Ishii \cite{Ishii70}. His method was also rather complicated. An essential advantage in his calculation was that he used the imaginary energies  $\epsilon_m=\pi T(2m+1)$ (also known as Matsubara frequencies or energies), where $m$ is an integer. This allowed him to write a single compact expression that included both bound state and continuum contributions. His general expression [Eq.\ (3.5) in  \cite{Ishii70}] can be written in the form \begin{eqnarray}
J=\frac{2e}{\hbar}T\sum_{m=-\infty}^\infty
\frac{\Delta^2\sin\phi}{(2\epsilon_m^2+\Delta^2)\cosh\frac{2\epsilon_mL}{\hbar v_{F}}+
2\epsilon_m\sqrt{\epsilon_m^2+\Delta^2}\sinh\frac{2\epsilon_mL}{\hbar v_{F}}+ \Delta^2\cos\phi}.
\label{e.ishimf}\end{eqnarray}
The same result has been obtained in several publications using different approaches \cite{Svidzinsky71,Svidzinsky73,Bezuglyi75,Kupriyanov81,Furusaki91,Bagwell92,Golubov04,Thuneberg23}. We comment on some of the calculations. Svidzinskii, Antsygina, and Bratus' solved the problem both by using the Gorkov equations and by using the Bololiubov-de Gennes equations, and discussed the problem in the earlier work by Kulik \cite{Svidzinsky71,Svidzinsky73}. Bezuglyi, Kulik and Mitsai already went on to study more general problems, but commented on the earlier calculations and provided alternative expressions for the current in limiting cases \cite{Bezuglyi75}. Kulik and Omel'yanchuk used the quasiclassical Green's function satisfying the Eilenberger equation to study the special case $L=0$ \cite{Kulik77,Kulik78}. Kupriyanov used the same method in the case of general $L$ and derived the current in the same form as given in (\ref{e.ishimf}) \cite{Kupriyanov81}. He also discussed several limiting cases. A calculation aimed for conducting channels and  based on Bololiubov-de Gennes equations is given by Furusaki, Takayanagi and Tsukada \cite{Furusaki91}. A same type of derivation with plenty of explanations was given by Bagwell \cite{Bagwell92}. The derivation using the Eilenberger equation is presented in the review \cite{Golubov04} together with numerous studies of the Josephson current in various models. Recently the derivation of (\ref{e.ishimf}) has been presented once more  in response to false claims about its invalidity \cite{Thuneberg23}. In summary, formula (\ref{e.ishimf}) was discovered over 50 years ago, and it has formed the starting point for numerous other studies since.

Equation (\ref{e.ishimf}) is a highly convenient expression to calculate the current. The problem in (\ref{e.ishimf}) is its rather abstract form as a sum over imaginary energies. In order to gain physical understanding, one should do an analytic continuation of the imaginary energies $E=i\epsilon_m$ to the real $E$ axis.
We notice that the  denominator of (\ref{e.ishimf}) vanishes on the real axis at points $E=E_n$ satisfying 
\begin{eqnarray}
\phi(E)=\arccos\left[-(1-2\frac{E^2}{\Delta^2})\cos\frac{2E L}{\hbar v_{F}}+2\frac{E}{\Delta}\sqrt{1-\frac{E^2}{\Delta^2}}\sin\frac{2E L}{\hbar v_{F}}\right].
\label{e.phifeps}\end{eqnarray}
This is just another expression for the bound state energies given in (\ref{e.kedr2}).
We express the sum over $\epsilon_m$ as an integral and deform the integration path as explained in many textbooks, for example in Ref.\ \cite{FW}, and also in Ref.\ \cite{Ishii70}. The result is 
\begin{eqnarray}
J&=&\frac{2e}{\hbar}\sum_{n}\frac{dE_n}{d\phi}f(E_n)
+\frac{2e}{\hbar}
\left(\int_{-\infty}^{-\Delta}+\int_ \Delta^\infty\right)\nonumber\\&&\times
\frac{\frac2{\pi}|E|\sqrt{E^2-\Delta^2}\sin\frac{2E L}{\hbar v_{F}}\Delta^2\sin\phi}{[(\Delta^2-2E^2)\cos\frac{2E L}{\hbar v_{F}}+\Delta^2\cos\phi]^2+4E^2(E^2-\Delta^2)\sin^2\frac{2E L}{\hbar v_{F}}}f(E)dE.
\label{e.Jcont}\end{eqnarray}
The first term is the contribution from  the bound states at $|E|<\Delta$. It is the sum over the energy values satisfying (\ref{e.phifeps}). The second term is the contribution from  the branch cut of the square root in the continuum $|E|>\Delta$. In both terms a multiplier is the Fermi function $f(E)=1/(e^{E/T}+1)$. The derivation given here proves formula (\ref{e.Jcont}) for equilibrium occupations of the levels, but the equation is valid also for distributions $f(E)$ differing from $1/(e^{E/T}+1)$.

A noteworthy feature of the expression (\ref{e.Jcont}) is that it is a sum and integral over both positive and negative energies. This is a consequence of the Green's function technique, where a hole-type excitation (a particle is removed from the ground state) appears as a negative-energy pole and a particle-type excitation (a particle added to the ground state) as a positive-energy pole. We call the presentation of both negative and positive energies the {\em semiconductor picture}, since it makes the superconducting gap look like the gap between the valence and conduction bands in a semiconductor. An alternative to the semiconductor picture is the {\em excitation picture}, which shows only the excitation energies, which are positive for both types of excitations. This is used in Fig.\ \ref{f.pot}. In the following, we prefer the excitation picture for two reasons. Firstly, consider a level of a given wave vector that decreases in energy from positive to negative. In the semiconductor picture this goes smoothly, in contrast to the excitation picture, where an excitation of the given momentum disappears and an excitation of the opposite momentum is created. Secondly, the equilibrium current (the supercurrent) can be interpreted to arise mainly from filled negative energy levels since $f(E)$ in (\ref{e.Jcont}) is small for the positive energy levels (and vanishes at $T=0$). The semiconductor picture is a direct generalization of the simple model of the normal state, where the free electron levels are filled up to the Fermi energy.

\begin{figure}[tb] 
   \centering
   \includegraphics[width=0.32\linewidth]{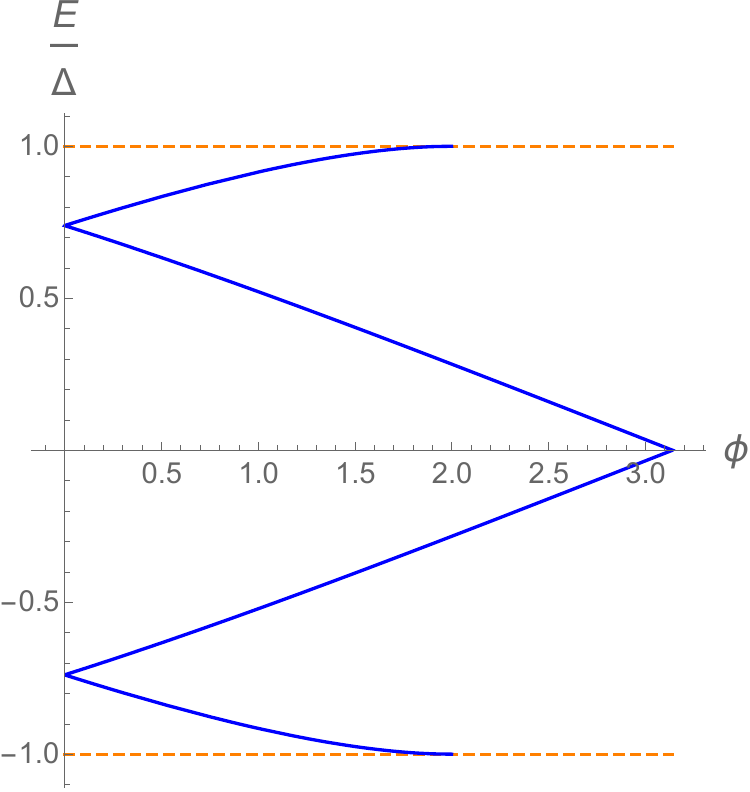} 
    \includegraphics[width=0.32\linewidth]{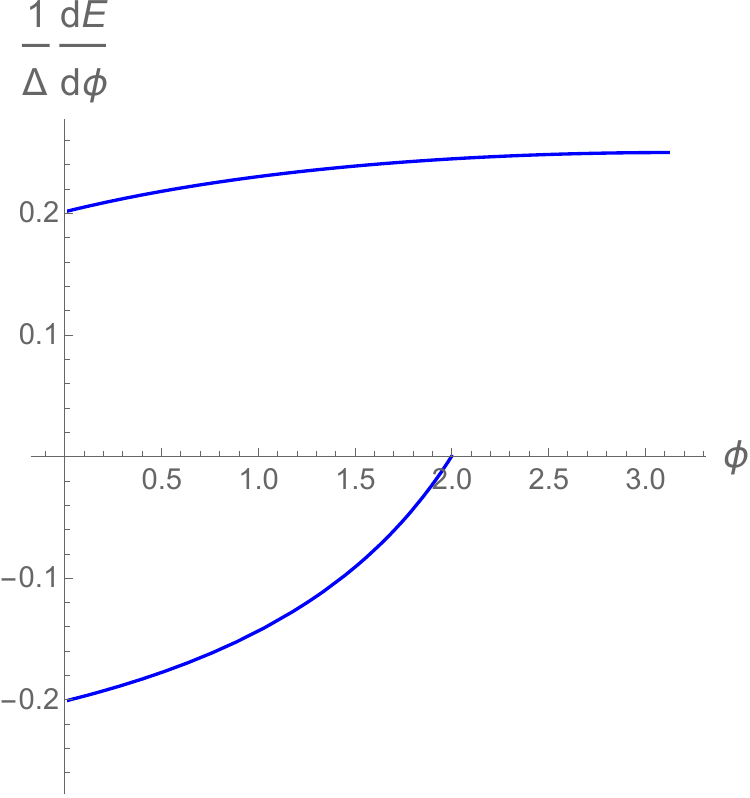} 
    \includegraphics[width=0.32\linewidth]{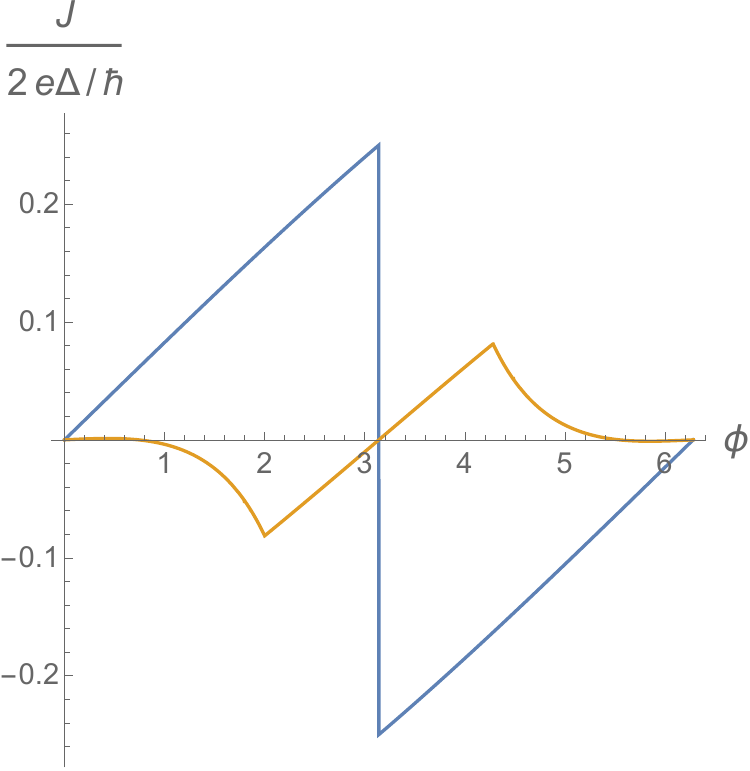}

   \caption{Numerical results in the square-well pair-potential model with a short N region. (left) The bound state energies as functions of the phase difference $\phi$ from  (\ref{e.phifeps}). (middle) The contributions of the filled bound states ($E<0$) to the current are given by the derivatives of the bound state energies (\ref{e.Jcont}). (right) The total current  (blue) (\ref{e.ishimf}) and the contribution of the continuum (orange) [integral term in (\ref{e.Jcont})].
   The parameters are $L\Delta/\hbar v_F=1$ and  $T=0$.}
   \label{f.L1}
\end{figure}

\begin{figure}[tb] 
   \centering
   \includegraphics[width=0.32\linewidth]{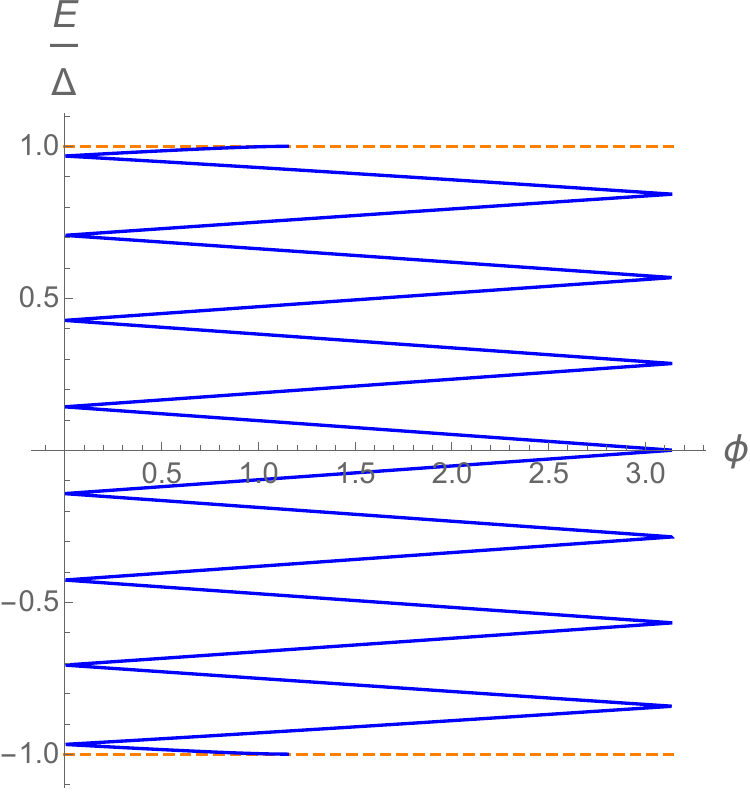} 
    \includegraphics[width=0.32\linewidth]{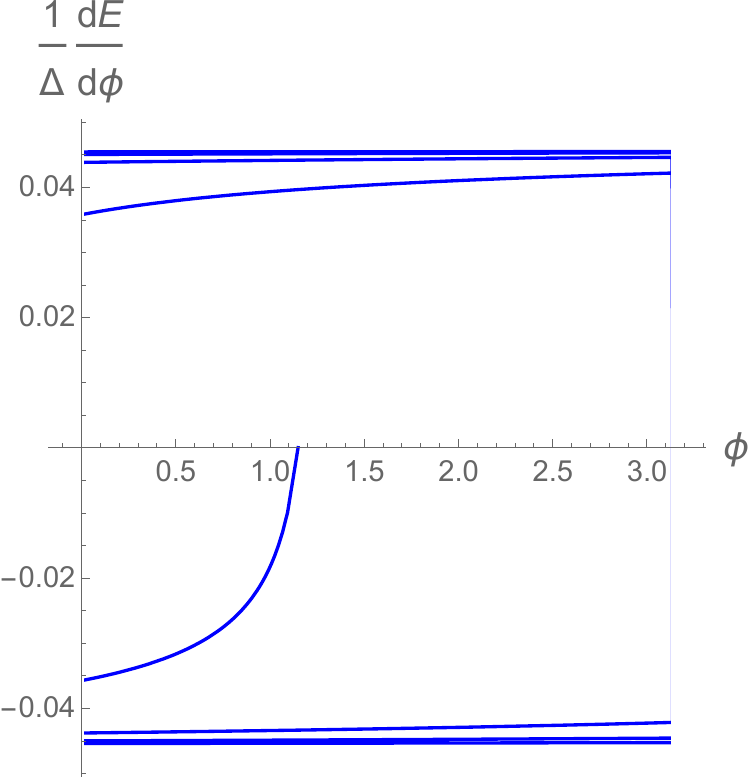} 
     \includegraphics[width=0.32\linewidth]{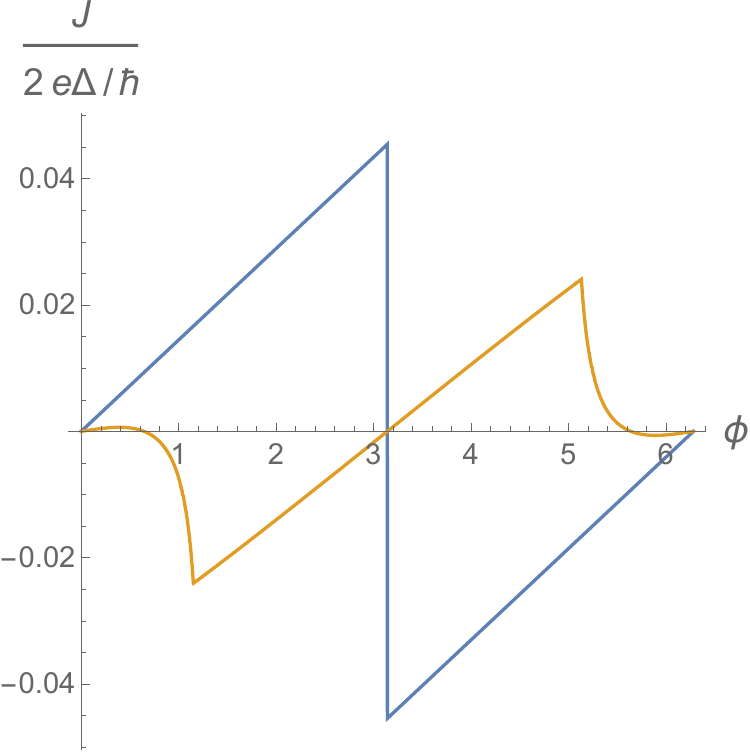} 
    \caption{Same as Fig.\ \ref{f.L1} but for  a longer N region, $L\Delta/\hbar v_F=10$. There are more bound states. Their currents alternate between values $\pm 0.045$ approximately, except the energy level that merges with the continuum at $\phi_c=1.150$.}
   \label{f.L10}
\end{figure}

\begin{figure}[tb] 
   \centering
   \includegraphics[width=0.32\linewidth]{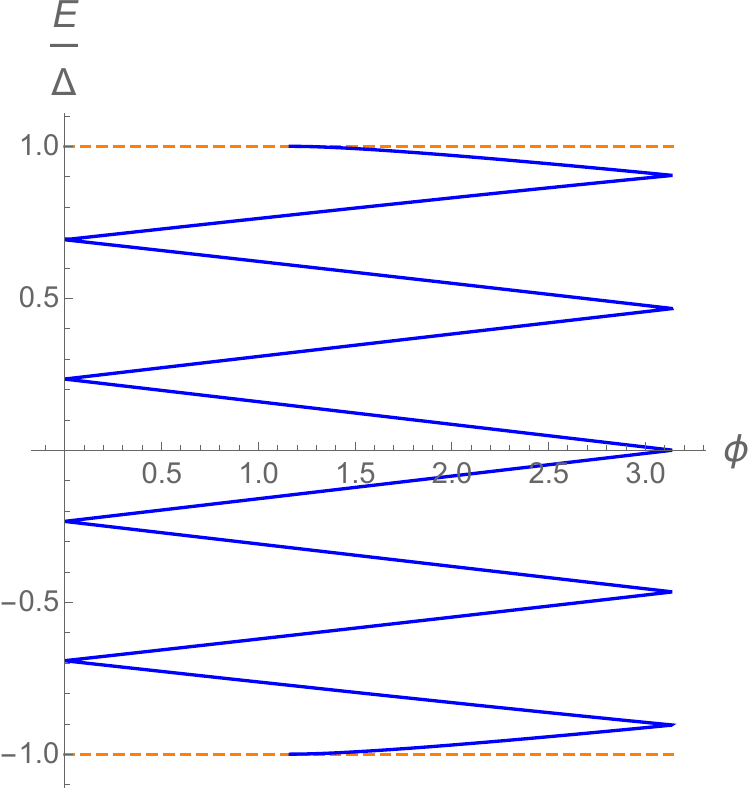} 
    \includegraphics[width=0.32\linewidth]{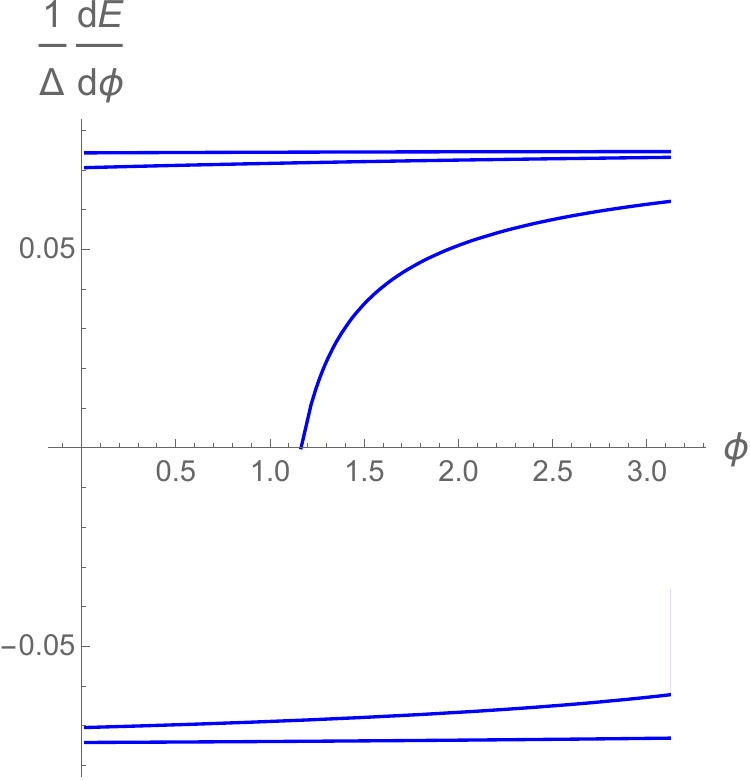} 
     \includegraphics[width=0.32\linewidth]{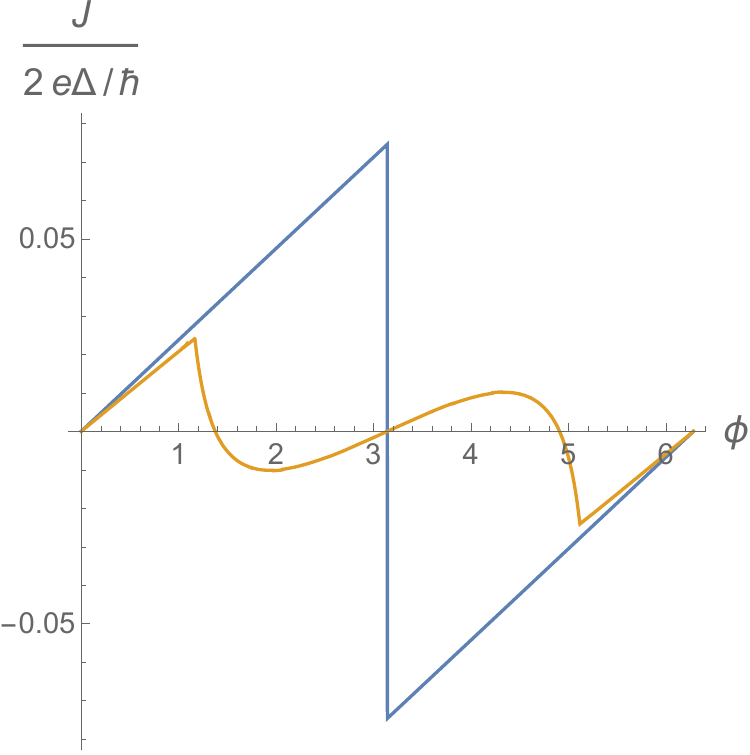} 
    \caption{Same as Figs.\ \ref{f.L1} and \ref{f.L10} but $L\Delta/\hbar v_F=5.7$. This is an example of a bound state emerging from the continuum with increasing $\phi$.}
   \label{f.L57}
\end{figure}

We study expression (\ref{e.Jcont}) for  the current by evaluating the different terms numerically at zero temperature, $T=0$. The results are presented in Figures \ref{f.L1}-\ref{f.L57}. The difference between the three figures is the length of the normal region.
The left panels in the figures show the energies of the bound states as functions of the phase difference $\phi$. These are conveniently obtained by plotting $\phi$ as a function of $E$ (\ref{e.phifeps}) and then exchanging the axes. Because of the symmetry of the energy levels with respect to $E=0$, we discuss only the levels at negative energies, which are filled in equilibrium at $T=0$. With increasing $\phi$ the levels degenerate at $\phi=0$ shift up or down in energy depending on their direction of momentum, as illustrated in Ref.\ \cite{Thuneberg23}.
In the case of Fig.\ \ref{f.L1}, there are two bound states for small $\phi$. At a critical $\phi_c=2$ the negative-momentum level merges with the continuum, so that  only the positive-momentum bound state survives in the range $\phi_c<\phi<\pi$. 
 
The contribution of a bound state to the current (\ref{e.Jcont}) is given by the derivative of its energy with respect to the phase. These are evaluated in the middle panels of Figs.\ \ref{f.L1}-\ref{f.L57}.
In the case of Fig.\ \ref{f.L1}, we see that the currents of the two filled bound states are in opposite directions. Their currents  cancel at $\phi=0$. With increasing $\phi$ the negative contribution decreases leaving a net positive current (in units of $2e\Delta/\hbar$, where $e$ is the electron charge). At $\phi>\phi_c$ the negative-momentum level has moved to the continuum so that the bound-state current comes from the single positive-momentum level. 

The right panels  of Figs.\ \ref{f.L1}-\ref{f.L57} show the continuum current (\ref{e.Jcont}) by orange lines. We see that it vanishes only at special points $\phi=n\pi$ with integer $n$. In the case of Fig.\ \ref{f.L1}, it has a maximal negative value at the cusp at $\phi=\phi_c$, where the negative-momentum bound state disappears. The total current is given by the blue line. This equals the sum of the bound-state currents (middle panel) and the continuum contribution (orange line). We see that the cusp in the continuum current just cancels the vanishing of bound-state current of the negative-momentum level at  $\phi_c$. The total current does not exhibit any special behavior at $\phi_c$. Similar conclusions are obtained in all Figs.\ \ref{f.L1}-\ref{f.L57}. The insensitivity of the current on the bound states near the gap edge was related to scattering theory in Refs.\ \cite{Svidzinsky71,Svidzinsky73}. It can also be argued that the current is determined by levels near the Fermi level and thus should be independent of the details at the relatively high energy of the gap edge \cite{Thuneberg23}. Further plots of the different contributions to the current can be found in Refs.\ \cite{Bagwell92,Gunsenheimer94}.

We see from the plots above that the zero-temperature current is discontinuous at $\phi= \pi$. The current in the neighborhood of  this point is 
\begin{eqnarray}
 J(\phi=\pi\mp0^+, T=0)=\pm\frac{ev_F}{L^*},\quad L^*=L+\frac{\hbar v_F}{\Delta}.
\end{eqnarray}
This current comes solely from the two levels that cross the Fermi level at $\phi=\pi$, as the contributions from all other bound states cancel each other and the continuum contribution vanishes. The quantity $L^*$ can be interpreted as the effective length of a bound state, where the latter term in $L^*$ describes the penetration of the bound state in the superconducting banks \cite{Bardeen72}.

The square-well pair-potential model has a few limiting cases. The case of the {\em zero-length potential well}, $L=0$, was first studied by
Kulik and Omel'yanchuk \cite{Kulik77,Kulik78}. It is easy to calculate from equations (\ref{e.phifeps}) and (\ref{e.Jcont}) the bound state energies $E=\pm\Delta\cos(\phi/2)$ and the current
\begin{equation}
J(L=0)=\frac{e}{\hbar}\Delta\sin\frac\phi2\tanh\frac{\Delta\cos(\phi/2)}{2T}.
\label{e.bardeen3}\end{equation}
The current comes from the two bound states only as the continuum contribution vanishes.

The second limiting case we consider is the {\em infinite potential well}. For an ordinary potential, this model is well presented in books of quantum mechanics because it has a fully analytic solution. In the case of the pair potential we take the limit $\Delta\rightarrow \infty$.  It is easy to calculate from equations (\ref{e.phifeps}) and (\ref{e.ishimf}) that the bound states have energies
$E_n=\frac{\hbar v_{F}}{2L}[\pm \phi+(2n+1) \pi]$ and the current \cite{Bezuglyi75}
\begin{eqnarray}
J(\Delta\rightarrow\infty)=\frac{2e}{\hbar}T\sum_{m=-\infty}^\infty\frac{\sin\phi}{\cosh(2\epsilon_mL/\hbar v_{F})+ \cos\phi}.
\label{e.ishimfdi}\end{eqnarray}
The infinitely deep potential well is a good approximation for low-energy levels, $E_n\ll\Delta$ in a long junction, $L\gg\hbar v_F/\Delta$ because in this limit the bound state energies (\ref{e.kedr2}) become independent of $\Delta$. This model is equivalent to the one used by Bardeen and Johnson \cite{Bardeen72} (except that they use $L^*$ instead of $L$).

We study the current (\ref{e.ishimfdi}). At $T=0$ the sum goes to an integral. Evaluating this gives
\begin{eqnarray}
J(\Delta\rightarrow\infty,T=0)=\frac{ev_{F}}{\pi L}\phi \quad {\rm for}\ |\phi|<\pi
\label{e.ishimfd0}\end{eqnarray}
and repeated periodically with period $2\pi$. This result was found by Ishii (the linear dependence on  $\phi$ in  Ref.\ \cite{Ishii70} and the coefficient in Ref.\ \cite{Ishii72}). We see that the linear dependence is rather well satisfied in the selected examples  of the general model, Figs.\  \ref{f.L1}-\ref{f.L57}.

An interesting feature of the current (\ref{e.ishimfd0}) is that it corresponds to motion at a velocity $v_s=(\hbar/2m)\phi/L$, which can be interpreted as the superfluid velocity. Thus the normal region seems not to differ from a superconductor, as long as the critical velocity $v_c=\pi \hbar/2mL$ is not exceeded. Expressed in other words, the system behaves like it would be Galilean invariant, although it is not because the locations of the N-S interfaces are fixed \cite{Bardeen72}.  

\begin{figure}[tb] 
   \centering
     \includegraphics[width=0.32\linewidth]{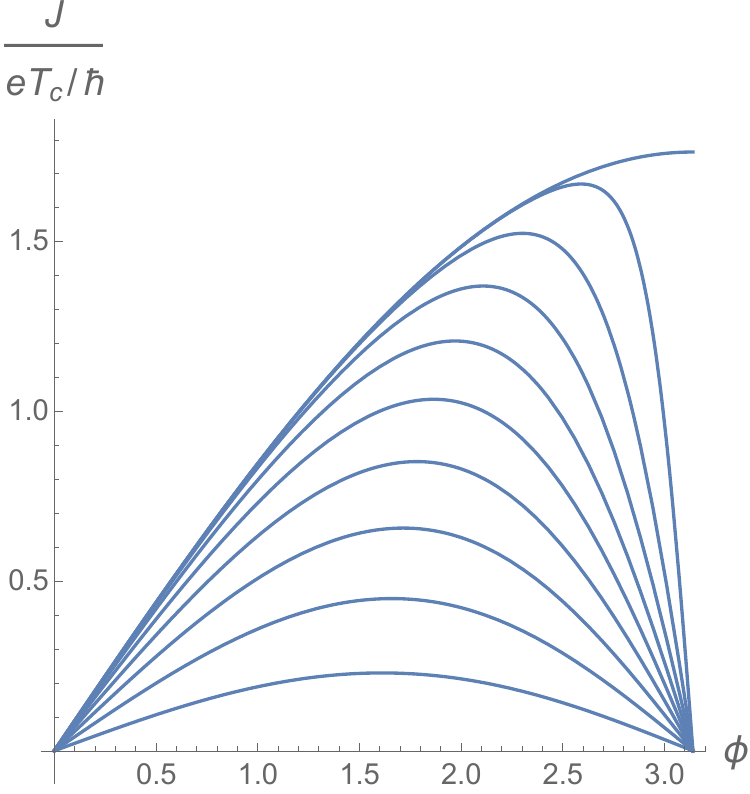}
   \includegraphics[width=0.32\linewidth]{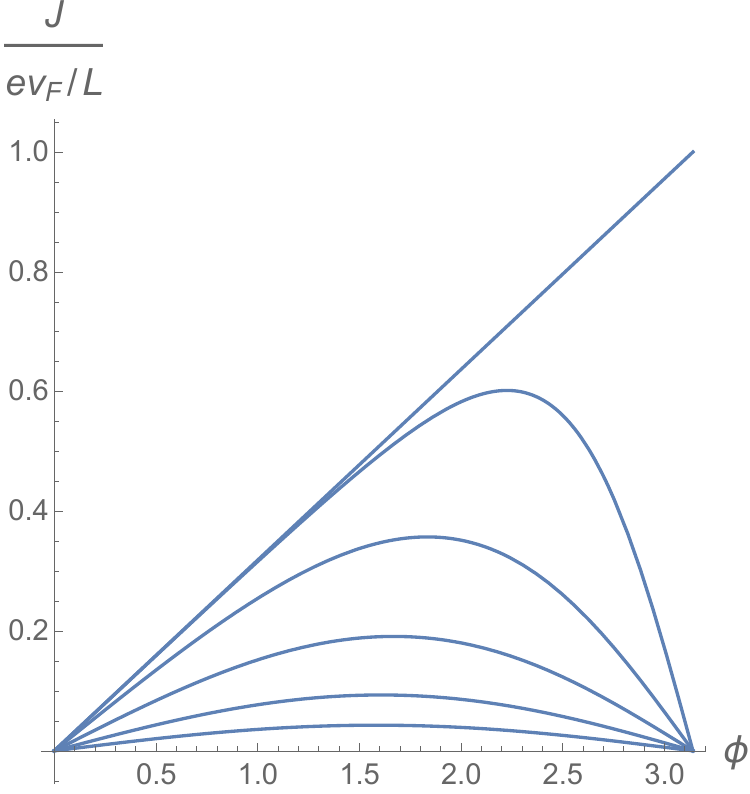} 
    \caption{The current-phase relations at different temperatures. (left) Zero-length pair-potential well (\ref{e.bardeen3}) evaluated using the BCS gap function $\Delta(T)$ at reduced temperatures $T/T_c= 0, 0.1, \ldots, 0.9$.
  (right)  Infinitely-deep pair-potential well (\ref{e.ishimfdi}) evaluated at temperatures 0, 1,\ldots , 5 in units of $\hbar v_F/2\pi L$. In both panels, an increasing temperature corresponds to lower current, but the scales of the current and the temperature differ essentially. 
  }
   \label{f.td}
\end{figure}

The temperature dependence of the current can be calculated from (\ref{e.ishimf}) in the general case.  Examples in two limiting cases are shown in Figure \ref{f.td}.

Let us consider the application of the square-well pair-potential model to experimentally realizable systems. 
The main deficiency of the model in several applications is that the pair-potential is not self-consistent. For example, consider a normal section in a uniform-thickness superconducting wire. Because ordinary scattering was neglected, the normal region induces a proximity effect on the superconducting banks. As a result, the pair potential is reduced in the banks near the normal region. This affects especially the levels near the gap edge, which can penetrate deeper into the banks.
There are, however, some cases where the square-well model can be justified quantitatively.

\begin{itemize}
\item The limit of a long junction, $L\gg\hbar v_F/\Delta$. Here the current is determined by low-energy levels, $E_n\ll\Delta$. The energies (\ref{e.kedr2}) of these levels are not dependent on $\Delta$. Thus we expect this to be the case also for $\Delta(x)$ modified by the proximity effect. 

\item The case of a weak link separating two wider superconducting leads. In this case the order parameter in the banks is determined by quasiparticle trajectories staying in the leads with a minimal contribution from the transmitting trajectories \cite{Kulik77}.

\item The case that the Fermi velocity in the banks is much larger than in the normal region. This means that only a few of the trajectories incident on the junction will be transmitted, and thus the proximity effect on the banks is small \cite{Kupriyanov81}.

\end{itemize}

Another possible point of worry in the square-well pair-potential model could be that current conservation is strictly valid only in the normal region. There is no current deep  in the superconducting regions of constant phase. This problem can be corrected by adding a spatially varying phase in the superconducting regions. In the three cases where the proximity effect in the banks can be neglected, the current in the normal region is always small compared to the critical current in the superconducting regions. Thus the induced phase gradient in the superconducting regions is small and does not affect the calculation of the  bound states and the current. This means that current conservation is not a problem that is separate from the self-consistency problem. Thus attempts to restore current conservation without solving the self-consistency of the pair potential are misguided. 

The square-well pair-potential model can be used as a qualitative model in cases where  its assumptions are not strictly valid.
For example, consider a flux line in a type II superconductor and  a quasiparticle trajectory passing the vortex line at distance $b$ (impact parameter). There is phase change along the trajectory. We can model this using the square-well model. The phase along the trajectory changes by $2\pi$ as a function of $b$ changing from $-\infty$ to $+\infty$. There has to be a bound state whose energy changes from positive to negative energy, and an opposite momentum state that shifts from negative to positive energy. The zero crossing takes place as the trajectory goes through the vortex line, $b=0$. This gives a qualitative semiclassical description of the Caroli-de Gennes-Matricon vortex-core states.

The model can also be used to demonstrate the difference between an ideal normal metal and a superconductor. Consider an occupation of the levels that the positive momentum levels are filled up to an energy $E>0$ and correspondingly, there is a deficit in filling the negative momentum levels. Such a state would not decay if we strictly assumed the pure limit. The current in such a state can be considerably higher than that obtained with equilibrium occupation. This is not supercurrent since, for a filled positive momentum level, there are empty negative momentum levels at lower energy. Inelastic scattering can reduce the distribution and the current to their equilibrium values.

\begin{figure}[tb] 
   \centering
   \includegraphics[width=0.32\linewidth]{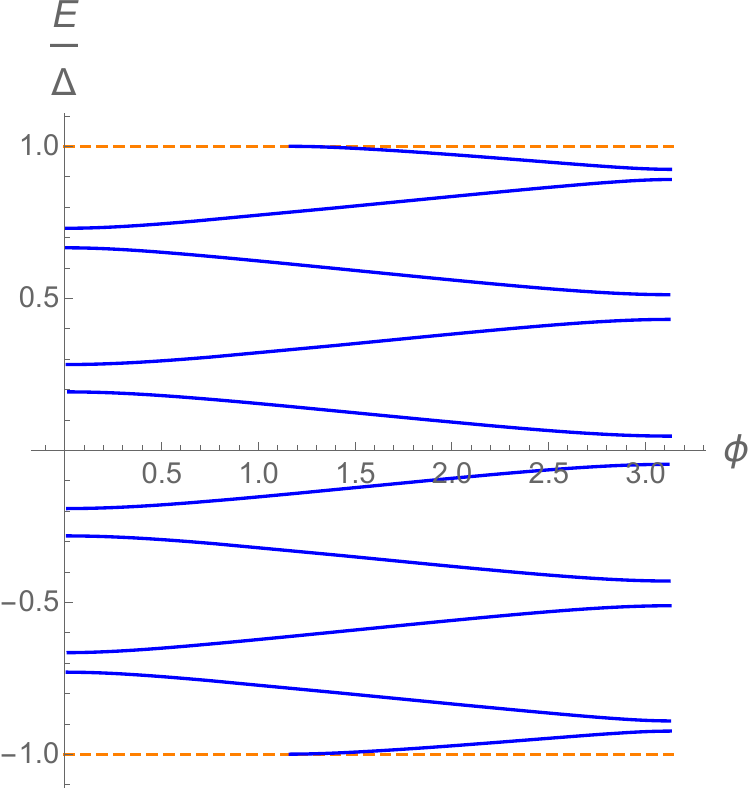} 
    \includegraphics[width=0.32\linewidth]{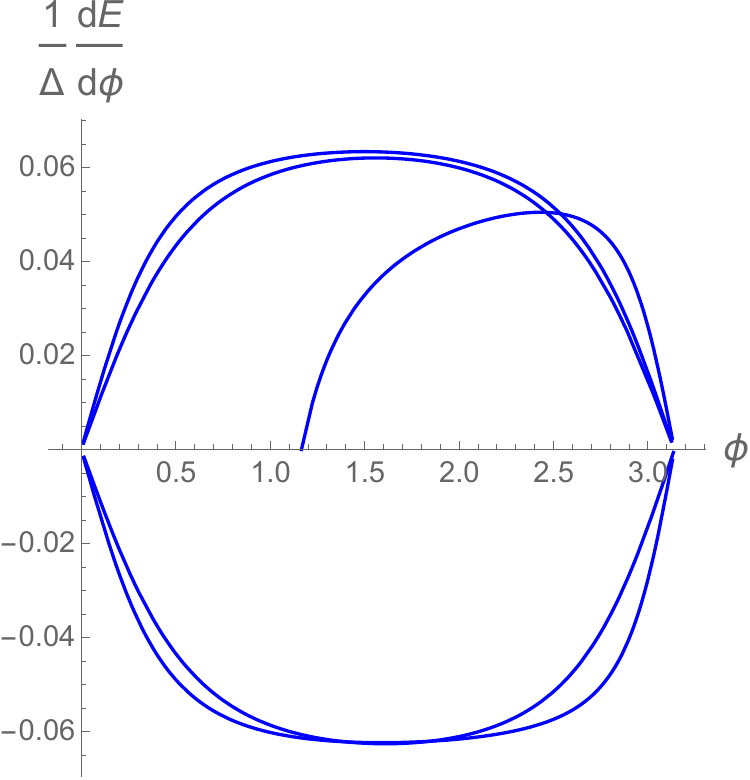} 
     \includegraphics[width=0.32\linewidth]{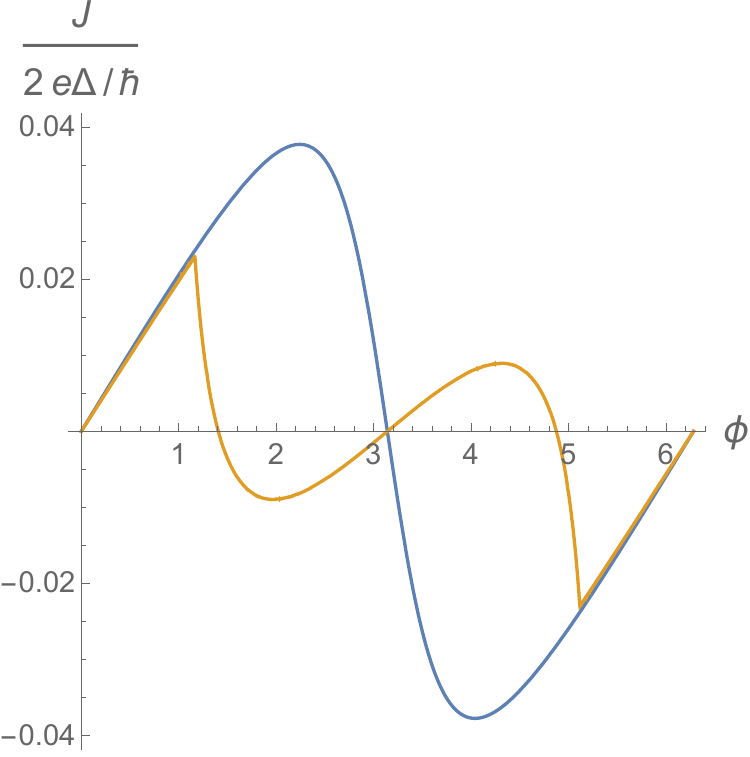} 
    \caption{Same as Fig.\ \ref{f.L57}  but including barriers at both NS interfaces. The noteworthy difference to Fig.\ \ref{f.L57}  is the opening of gaps in the energy spectrum at $\phi=0$ and $\pi$. The figure is based on the calculation  by Galaktionov and Zaikin \cite{Galaktionov02}. The parameters are the transmission probability 0.9 at both barriers, $L\Delta/\hbar v_F=5.7$, zero temperature, and the phase acquired in one round trip between the two barriers equal to $\pi/2$. [Note the misprint  of an extra factor $\sin\chi$ in formula (16) of  Ref.\ \cite{Galaktionov02}.] }
   \label{f.L57B}
\end{figure}

There are a great  number of generalizations of the square-well pair-potential model. The most obvious is to allow for a self-consistent pair potential, see Refs.\ \cite{LevyYeati95,Riedel96} for some early calculations. A second important generalization is to consider the normal scattering of electrons. This couples the levels with opposite momentum and results in that  the crossings of the opposite momentum levels at $\phi=n\pi$ (integer $n$) are replaced by avoided crossings \cite{Bezuglyi75,Haberkorn78,Beenakker91,Bagwell92,Galaktionov02}. This is illustrated in Fig.\ \ref{f.L57B}.
There are other generalizations to non-equilibrium phenomena, different geometries, unconventional superconductors/superfluids etc. Some of the topics are considered in Refs.\ \cite{Golubov04,Belzig99,ABS18}. In fact, all studies of inhomogeneous superconductors can be seen as generalizations of the square-well pair-potential model. 

I will finish this article with a personal view of how I learned and used the square-well pair-potential model. First, an impurity in the vortex core acts similarly as in a SNS junction at phase difference $\phi=\pi$: it increases the superconductor's condensation energy. This leads to a pinning force on the vortex to the impurity \cite{TKR84,T84}. This is a single-impurity version of the effect known earlier that impurity scattering makes the superconducting coherence length shorter.
Second, an impurity or a wall in a p-wave superfluid acts like a SNS junction as the phase on the quasiparticle trajectory jumps  in scattering \cite{TKR81,Zhang87,Hanninen03}.
Third, we have studied both equilibrium and non-equilibrium currents in pinholes, the analog of point contacts in superconductors, of superfluid $^3$He \cite{Viljas02,Viljas04,Thuneberg06}. This allows one to explain $\pi$ states observed experimentally in an array of pinholes.


In summary, similarly to the role of the square-well potential model to learn quantum mechanics, the square-well 
pair-potential model is an analytically solvable model that allows us to learn inhomogeneous superconductivity and fermion superfluidity.


\end{document}